\documentstyle[preprint,aps]{revtex}
\begin{document}
\preprint {WIS-96/17/Apr.-PH} 
\draft
\title{Scaling approach to higher-twist corrections}
\author{S.A. Gurvitz$,^{(1)}\;$A. Mair,$^{(2)}\;$ and M. Traini$.^{(2)}\;$}
\address{$^{(1)}$Department of Particle Physics, Weizmann Institute of
Science, Rehovot 76100, Israel\\
$^{(2)}$Dipartimento di Fisica, Universit\`a degli Studi di Trento
and Istituto Nazionale di Fisica Nucleare, G.C. Trento, I-38050 Povo,
Trento, Italy}
\maketitle
\begin{abstract}
Bjorken scaling is violetd. At large $x$-values ($0.7 \lesssim x \leq 1$) the
violation is mainly attributed to $\propto 1/Q^2$ (higher-twist) corrections.
We discuss how to incorporate such corrections by using a new scaling variable 
$\bar x(x,Q^2)$ which accounts for the non-perturbative effects due to the
confining parton interaction in the final state. The approach can accomodate
also the remaining QCD logarithmic corrections and the behaviour of the
$F_2(x,Q^2)$ structure functions is reproduced, in a quite natural way, within 
a wide kinematical range.
\end{abstract}
\vskip 1cm
\par
The existing data for hadron structure functions, $F_2(x,Q^2)$, 
show considerable $Q^2$-dependence, which is mainly attributed 
to the QCD logarithmic corrections to Bjorken scaling. However, 
at $x\to 1$ the scaling violations are dominated by power corrections 
$\propto 1/Q^2$ (higher twist and target mass effects):
\begin{equation}
F_2(x,Q^2)= F_2^{as}(x,Q^2)+\frac{B(x)}{Q^2}+\cdots,
\label{a1}
\end{equation}
where $F_2^{as}(x,Q^2)=F_2(x,Q^2\gg |B(x)|)$ and the remaining
$Q^2$-dependence in $F_2^{as}(x,Q^2)$ is to be attributed to QCD logarithmic 
corrections only. 

Power corrections can be incorporated in the first term of Eq.~(\ref{a1}) by 
using a different scaling variable, 
\begin{equation}
\hat x =\phi (x,Q^2)=x+\frac{b(x)}{Q^2}+\cdots,
\label{a3}
\end{equation}  
so that
\begin{equation}
F_2(x,Q^2)=F_2\left (\phi^{-1}(\hat x,Q^2),Q^2\right )
\simeq F_2^{as}(\hat x,Q^2)\,. 
\label{a2}
\end{equation}
The coefficient $B$, which determines the value of the power 
correction in Eq.~(\ref{a1}), is thus related to the structure function by  
$B(x)=b(x)\, \partial F_2^{as}(x,Q^2)/\partial x$. 

Actually, an analysis of data 
in terms of an appropriate scaling variable appears to be more convenient, 
than the direct evaluation of power corrections. For instance, it is  
common to use for an analysis of data the Nachtmann variable 
$\hat x\equiv\xi$\cite{nacht1}, 
\begin{equation}
\xi =\frac{2x}{1+\sqrt{1+4M^2x^2/Q^2}}\, ,
\label{a4}
\end{equation} 
which is expected to account for the effects of the target mass ($M$).

Besides the target mass effects, there are important nonperturbative 
effects from the confining interaction of the partons in the final state.  
Indeed, the partons are never free, so that
the system possesses a discrete spectrum in the final state.
Although in the Bjorken limit the struck quark can be considered 
a free particle, the discreteness of the spectrum manifests itself in 
power corrections to asymptotic structure functions\cite{greenb}.
One can anticipate that these corrections are significant in particular 
at large $x$, where lower-lying excitations should play an important role. 

A general analysis performed in the framework of Bethe-Salpeter equation 
shows that in the case of a local confining final state interaction 
the higher twist and target mass effects can  be effectively 
accounted for by taken the struck quark 
with the same off-shell mass before 
and after the virtual photon absorption\cite{gur}. As a result, the Bjorken 
scaling variable $x$ is replaced by a new scaling variable 
$\bar x\equiv\bar x(x,Q^2)$, which is   
the light-cone fraction of the {\em off-shell} struck quark.
Explicitly,
\begin{equation} 
\bar x=\frac{x+\sqrt{1+4M^2x^2/Q^2}-
\sqrt{(1-x)^2+4m_s^2x^2/Q^2}}
{1+\sqrt{1+ 4M^2x^2/Q^2}},
\label{a5}
\end{equation} 
where $M$ is the target mass and $m_s$ is the invariant mass of 
spectator partons (quarks and gluons). For $Q^2\to \infty$ or for $x\to 0$
the variable $\bar x$ coincides with the Nachtmann variable $\xi$, 
Eq. (\ref{a4}). However, at finite 
$Q^2$ these variables are quite different. 

It follows from Eq. (\ref{a5}) that $\bar x$ depends on the invariant
spectator mass, $m_s$. The latter can be considered a function 
of the external parameters only. In the limit $x\to 1$ 
(elastic scattering) no gluons are emitted, and thus $m_s\to m_0$, which is  
the mass of a two-quark system (diquark). When $x<1$, the spectator mass
$m_s$ increases due to gluon emission.   
For $x$ close to 1, $m_s$ can be 
approximated\cite{gur}
\begin{equation}  
m_s^2\simeq m_0^2+C(1-x), 
\label{a6}
\end{equation}
where the coefficient $C\sim$ (GeV)$^2$. In the following we regard  
it as a phenomenological parameter, determined from the data.

Let us consider the nucleon structure functions in the region of large $x$, 
where the power corrections to the scaling are dominant. 
At present, the only available large-$x$ data for proton and 
deuteron structure functions, $F_2^p(x,Q^2), F_2^d(x,Q^2)$, 
are the SLAC data\cite{slac1,slac2,slac3,slac4},
taken at moderate values of momentum transfer, $Q^2\lesssim 30$ (GeV/c)$^2$.
(The nucleon structure functions for higher values of momentum
transfer ($Q^2\lesssim 250$ (GeV/c)$^2$) are extracted 
from BCMDS\cite{bcd1} and NMC\cite{nmc1} data, yet only for $x\leq 0.75$). 
The SLAC data for the proton and deuteron structure functions  
for $x\geq 0.7$ and $5\lesssim Q^2\lesssim30$ (GeV/c)$^2$ are shown in 
Fig. 1 as a function of $x$. Also shown is the value of $F_2^p(x,Q^2)$ and 
$F_2^d(x,Q^2)$ for $Q^2$= 230 (GeV/c)$^2$ and $x=0.75$ taken from
the BCDMS data\cite{bcd1}. The data points close 
to the region of resonances were excluded by a requirement that the 
invariant mass of the final state  
$(M+\nu )^2-{\mbox{\boldmath $q$}}^2$ is greater than 
$(M+\Delta)^2$, where $\Delta=$ 300 MeV. In addition, we excluded the data  
points with $x>0.9$ from the deuteron structure only (Fig. 1b).     
The reason is that the deuteron structure function can not be 
represented as an average of the proton and neutron structure functions for 
$x\gtrsim 0.9$. Indeed, the calculations of Melnitchouk {\em et al.}\cite{mel}
show that the ratio  $2F_2^d/(F_2^p+F_2^n)$ is about 1.13 for 
$x=0.9$ and $Q^2$=5 (GeV/c)$^2$, and it rapidly increases for $x>0.9$. 
However, for $x<0.85$, this ratio is within 5\% of unity\cite{not1}.  
 
One finds from Fig.1 that the structure functions show 
no scaling in the Bjorken variable $x$. Also, very poor scaling is  
obtained when the data are plotted as a function of the Nachtmann 
variable $\xi$, Eq.~(\ref{a4})\cite{slac4}. However, the situation is 
different if we display the same data as 
a function of the variable $\bar x$, Eq.~(\ref{a5}).
It appears that the scaling in the $\bar x$-variable is strongly dependent 
on the value of diquark mass, $m_0$, Eq.~(\ref{a6}), but is much less 
sensitive to variation of the coefficient $C$. 
For instance, Fig. 2 shows the data as a function of $\bar x$ 
for $m_0$ = 600 MeV and $C$ = 3 (GeV)$^2$), i.e. by considering the spectator 
as build up from constituent quarks. The data display very poor scaling, 
although it is slightly better than that shown in Fig. 1. The scaling 
deteriorates even more when $m_0 > 600$ MeV.  
On the other hand, the scaling is very good both
for the proton and deuteron data, by taking
$m_0=0$, i.e. by considering the spectator build up by current
quarks\cite{gur}. The results are shown in Fig. 3, where the data are
plotted as a function of $\bar x$ for $m_0=0$ and $C$ = 3 (GeV)$^2$.
Note, that the high-$Q^2$ data points from BCDMS data\cite{bcd1} are
very close to the SLAC data points, taken at much lower values of $Q^2$.
Also note that $\bar x\to 1$ when $x\to 1$ for $m_0=0$, Eq.~(\ref{a5}) but
$\bar x(x,Q^2)<x$ for $x<1$. As a result the data, plotted as a function
of $\bar x$-variable are extended in a wider region, than the same data
plotted as a function of the $x$-variable (cf. Figs. 1 - 3).
 
Now by using the scaling variable $\bar x$, Eq.~(\ref{a5}) for $m_0=0$  
we are going to analyze the proton structure function for smaller values 
of $x$, where both power and logarithmic corrections 
to the Bjorken scaling play an important role. In this region  
($0.35 < x < 0.75$) the existing BCDMS\cite{bcd1} and NMC\cite{nmc1} data 
are extended up to much larger values of momentum transfer  
than the previously considered high-$x$  
SLAC data. It allows us to check our predictions in a wide $Q^2$ range.

QCD (logarithmic) evolution effects on $F_2$ are taken into account 
at Next-to-Leading Order (NLO)
\cite{nlo1} evolving {\em back}, in $Q^2$, the structure functions
starting from an asymptotic value of momentum transfer where the condition
$F_2(x,Q^2) \simeq F_2^{as}(x,Q^2)$ (cf. Eq.~(\ref{a1})) is 
fulfilled (in the
present case we choose $Q^2 = 230$ (GeV/c)$^2$, which is the highest value 
of the momentum transfer in the BCDMS data\cite{bcd1}). 
At that value of $Q^2$ the
functional form of $F_2$ is parametrized following the prescriptions of the
recent NMC fit \cite{nmcfit} and valid for a wide range of $x$ 
($0.006 < x < 0.9$). For the $x$-region we are interested in, namely 
$x \geq 0.35$,
one can assume that the {\em valence} contribution to $F_2$ dominates and
one can consider evolution of the nonsinglet (NS) components only. The 
accuracy of such an assumption can be deduced from the recent CTEQ 
parametrization \cite{cteq95} of the parton distributions.
The ratio $r= F_2^{NS}(x,Q^2)/F_2(x,Q^2)$ is $r\approx 0.96$ for $x=0.35$ and 
$r > 0.995$ for $x \geq 0.55$ and $Q^2 \geq 5$ (GeV/c)$^2$. 
In fact, since the only assumptions we are making
is due to the fact that the (small) singlet component evolves in $Q^2$ in a
different way with respect the dominant nonsinglet part, the inaccuracy
is less than $4\%$ for $x=0.35$ and less than $0.5\%$ for larger $x$.
For the actual calculations we use the NLO procedure developed in
ref. \cite{nlo2}. Under the renormalization group equation (RGE) the moments of
the nonsinglet components of the nucleon structure function,
$\langle F^{NS}(Q^2)\rangle_n = \int_0^1\,dx\,x^{n-2}\,F^{NS}(x,Q^2)$, evolve
according to 
\begin{equation}
\langle F_2^{NS}(Q^2)\rangle_n =  \langle F_2^{NS}(Q_0^2)\rangle_n\,\,
{1 + {\alpha_S(Q^2) \over 4 \pi}\,  R_{2,n}^{NS} \over 
 1 + {\alpha_S(Q_0^2) \over 4 \pi}\, R_{2,n}^{NS}}
\left( \alpha_S(Q^2) \over \alpha_S(Q_0^2) \right)^
{\gamma_{NS}^{0,n}/(2 \beta_0)}
\label{aa1}
\end{equation}
where 
\begin{equation}
R_{2,n}^{NS} = {\gamma_{NS}^{1,n} \over 2 \beta_0} - 
{\beta_1 \over 2 \beta_0^2}\,\gamma_{NS}^{0,n} + C_{2,q}^{1,n}
\label{aa2}
\end{equation}
and we used the expansion
\begin{equation}
{\alpha_S(Q^2) \over 4 \pi} =  {1 \over \beta_0 \,{\rm ln} (Q^2/\Lambda^2)}
- {\beta_1 \over \beta_0^3}\,{{\rm ln}{\rm ln} (Q^2/\Lambda^2) \over 
[{\rm ln} (Q^2/\Lambda^2)]^2}\, ,
\label{aa3}
\end{equation}
$\beta_0=11-2\,f/3$, $\beta_1=102-38\,f/3$ and $f$ is the number of
active flavours.
The values of all the other parameters of the calculation 
(like the anomalous dimensions $\gamma_{NS}^{0,n}$, $\gamma_{NS}^{1,n}$ and 
the Wilson coefficient $C_{2,q}^{1,n}$) have been compiled in 
refs.\cite{nlo1,thesis}.
The form (\ref{aa1}) guarantees complete symmetry for the evolution from 
$Q_0^2$ to $Q^2 > Q_0^2$ and {\em back}. In our case  
$Q_0^2=230$ (GeV/c)$^2$ and $5 < Q^2 < 230$ (GeV/c)$^2$. The NLO factorization
scheme implicitly selected in Eq.(\ref{aa2}) is the so called DIS scheme where
the structure function assumes the form $F_2^p(x,Q^2)=
\sum_q e^2_q\,x\,(q(x,Q^2) + \bar q(x,Q^2))$.

The procedure of perturbative QCD evolution previously sketched  
implies that the power corrections are already extracted from
the structure functions, so that the evolution has to be actually 
applied to $F_2^{as}$, Eqs.~(\ref{a1})-(\ref{a3})
\begin{equation}
F_2^{as}(x,Q^2)=F_2(x,Q^2)-\frac{B(x)}{Q^2}-\cdots\simeq
F_2\left (\bar x^{-1}(x,Q^2),Q^2\right ),
\label{a7}
\end{equation}
where we used the scaling variable $\phi (x,Q^2)\equiv \bar x(x,Q^2)$, 
Eq.~(\ref{a5}) for $m_0=0$ and $C$ = 3 (GeV)$^2$
that effectively incorporates the power corrections.
It implies that the evolution of structure functions 
{\em back} from $Q^2$ = 230 (GeV/c)$^2$, should be calculated with 
a shifted value of Bjorken variable, namely  
\begin{equation}
\Delta_1(x,Q^2)=F_2\left (\bar x^{-1}(x,Q^2),Q^2\right )-
F_2\left (\bar x^{-1}(x,Q^2),230\right )\, .
\label{a8}
\end{equation}
The influence of this shift would be quite important for low and moderated 
values of $Q^2$.
 
The corresponding variation of structure functions with $Q^2$ for 
fixed $x$ due to power corrections can be evaluated as   
\begin{equation}
\Delta_2(x,Q^2)=F_2\left (\bar x(x,Q^2),230\right )-
F_2\left (\bar x(x,230),230\right )\, .
\label{a9}
\end{equation}  
Finally the $Q^2$-dependence of structure functions due to logarithmic and 
power corrections to Bjorken scaling is given by
\begin{equation} 
F_2(x,Q^2)=F_2(x,230)+\Delta_1(x,Q^2)+\Delta_2(x,Q^2).
\label{a10}
\end{equation}
The results are shown in Figs. 4 and 5 for proton and deuteron structure 
functions respectively. The data points are from  SLAC and BCDMS data 
bins\cite{slac1,bcd1}. The dotted lines show the $Q^2$-dependence of the 
structure functions due to power corrections only, Eq.~(\ref{a9}).
The total $Q^2$-dependence of structure functions due to the power and
the logarithmic NLO corrections, Eq.~(\ref{a10}), is shown by the dashed 
and continuous lines for $\Lambda$ = 100 MeV and $\Lambda$ = 200 MeV
respectively. One finds from Figs. 4 and 5 that Eq.~(\ref{a10}) reproduces 
the experimental data in a large $Q^2$-range for both values of $\Lambda$, 
although the agreement is slightly better for $\Lambda$ = 100 MeV.  
In addition, since the results are strongly dependent on the spectator mass 
({\ref{a6}), it is remarkable that the same parameters $m_0=0$ and 
$C$ = 3 (GeV)$^2$ do reproduce the $Q^2$-behavior of the structure functions 
both for large and moderate $x$-values.

In conclusion we have presented a detailed investigation of the $F_2$
experimental data, both for proton and deuteron, analysed by means of the 
scaling variable $\bar x$ recently proposed in ref.\cite{gur}. 
Such variable includes the
non-perturbative effects due to the confining interactions of the partons in
the final state and contains, in an effective way, higher-twist corrections.
These contributions show up at large $x$ and at low and moderate momentum
transfer $Q^2$. The scaling behavior of the experimental data is improved in a
wide range of $x$ and $Q^2$. The additional inclusion of the QCD radiative
logarithmic corrections allows us the investigation of the structure function
in an even larger kinematical range with the extra bonus of a rather precise
identification of the spectator diquark mass.

\section{Acknowledgments}

We are grateful to A. Bodek and S. Rock for providing us with 
data files for proton and deuteron structure functions.

\newpage

\centerline{FIGURE CAPTIONS}

\vspace{20mm}

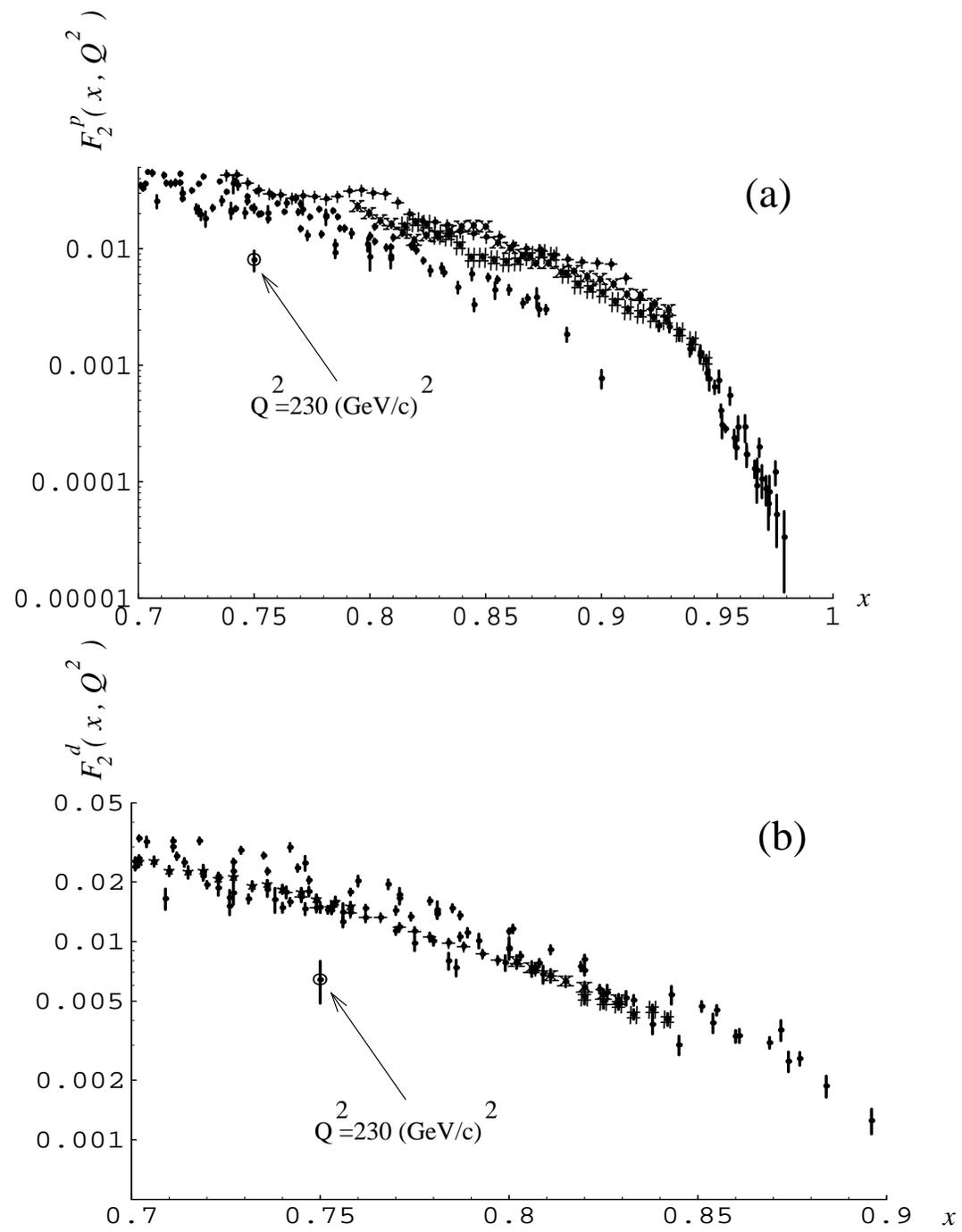
\noindent {\bf Fig 1.} {The SLAC data\cite{slac1,slac2,slac3,slac4}  
($5\lesssim Q^2\lesssim30$ (GeV/c)$^2$)
for proton (a) and deuteron (b), are shown as a function of the Bjorken variablle
$x$. Three high-statistics data sets\cite{slac3}  
for $Q^2\simeq$5.7, 7.6, and 9.5 (GeV/c)$^2$ are marked by  
``+", ``x", and ``{\#}" respectively.
The point at $Q^2=230$ (GeV/c)$^2$ and $x=0.75$ is from ref.\cite{bcd1}.}

\vspace{10mm}

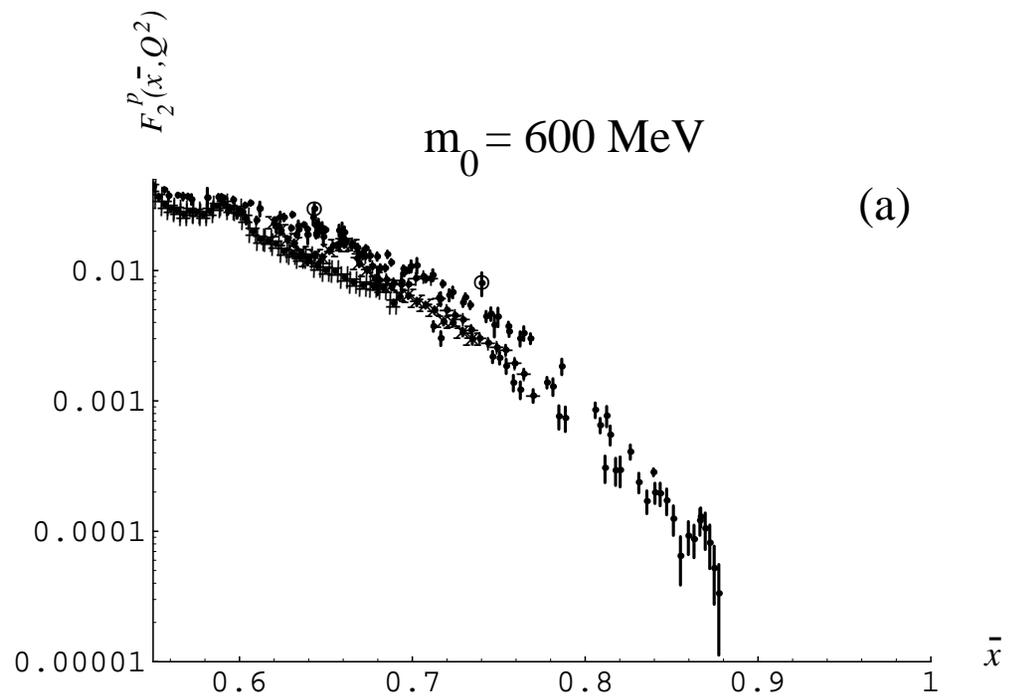
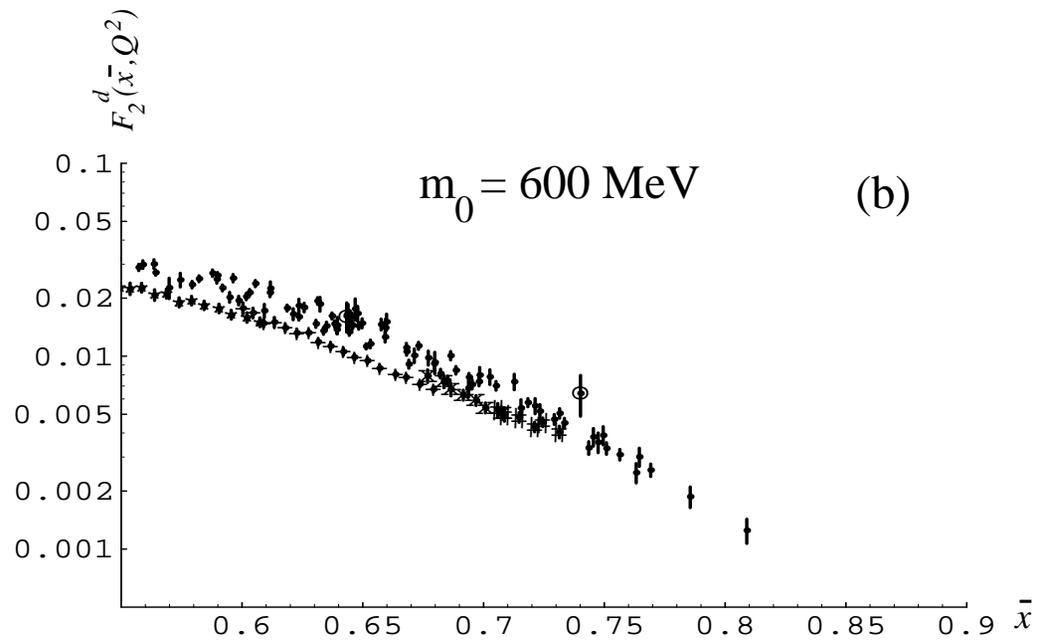
\noindent {\bf Fig 2.} {The data of Fig. 1 are shown as function of the 
$\bar x(x,Q^2)$ --- the scaling variable of Eq.~(\ref{a5}) --- 
assuming  $m_0=600$ MeV and $C$ = 3(GeV)$^2$ for the spectator mass $m_s$, 
Eq.~(\ref{a6}).}

\vspace{10mm}

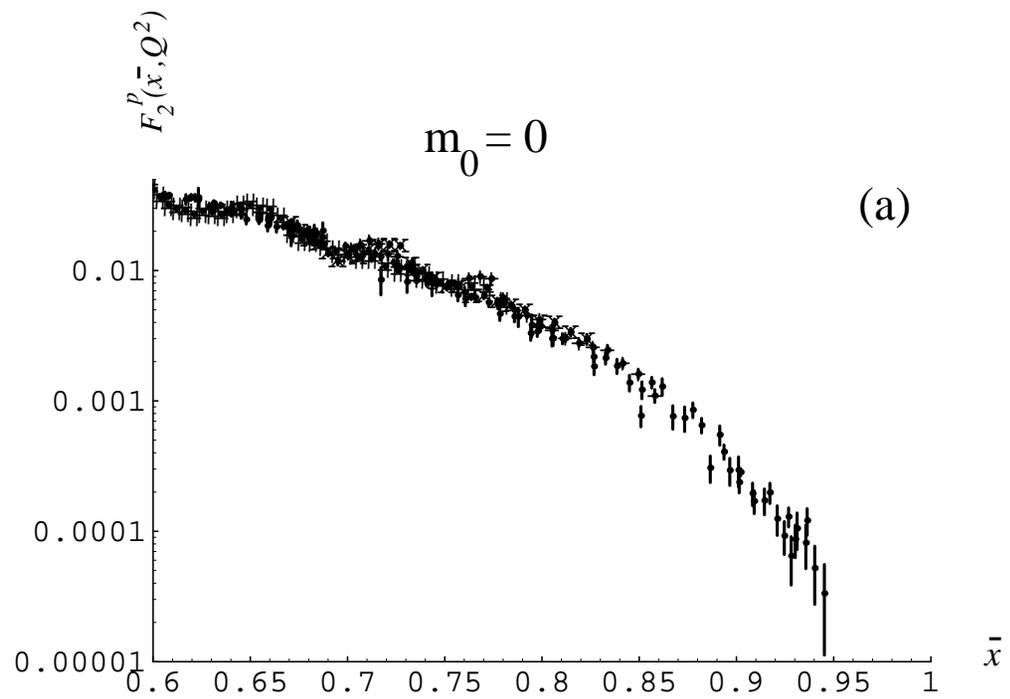
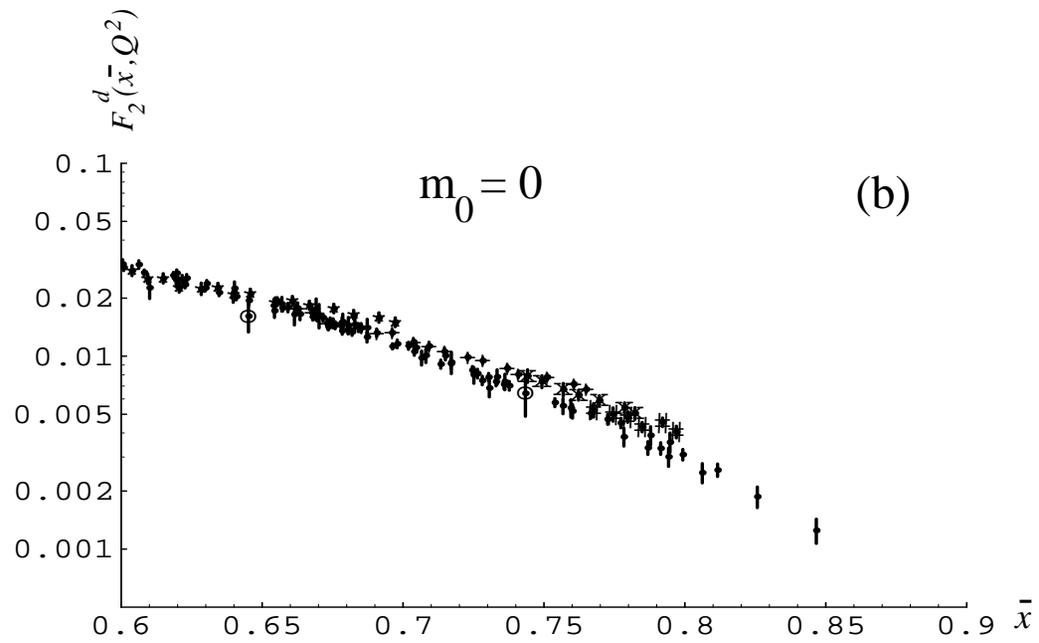
\noindent {\bf Fig 3.} {As in Fig. 2 assuming $m_0=0$
and $C$ = 3 (GeV)$^2$.}

\vspace{10mm}

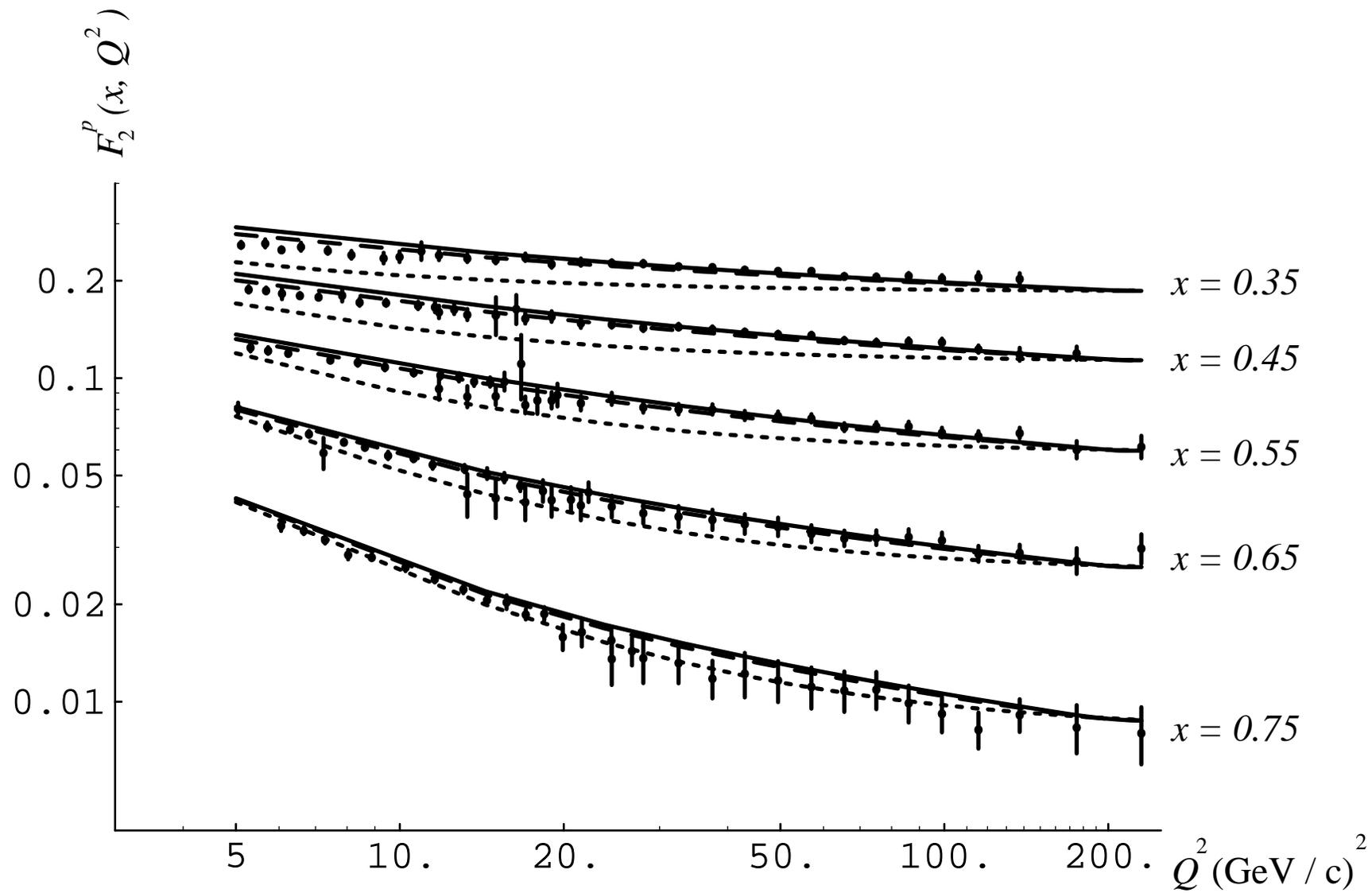
\noindent {\bf Fig 4.} {The proton structure function $F_2^p(x,Q^2)$ is shown
as a function of $Q^2$ at different $x$-values. The dotted lines include power
corrections only. They are evaluated according to Eq.~(\ref{a9}) and the
scaling variable $\bar x$ of Eqs.~(\ref{a5}), (\ref{a6}) with $m_0=0$
and $C$ = 3 (GeV)$^2$. The additional QCD logarithmic corrections evaluated at
NLO according to the procedure of Eq.~(\ref{a8}), (\ref{a10}) for different 
$\Lambda$ scales are shown by the dashed ($\Lambda = 100$ MeV) and 
continuous lines ($\Lambda = 200$ MeV).}

\vspace{10mm}

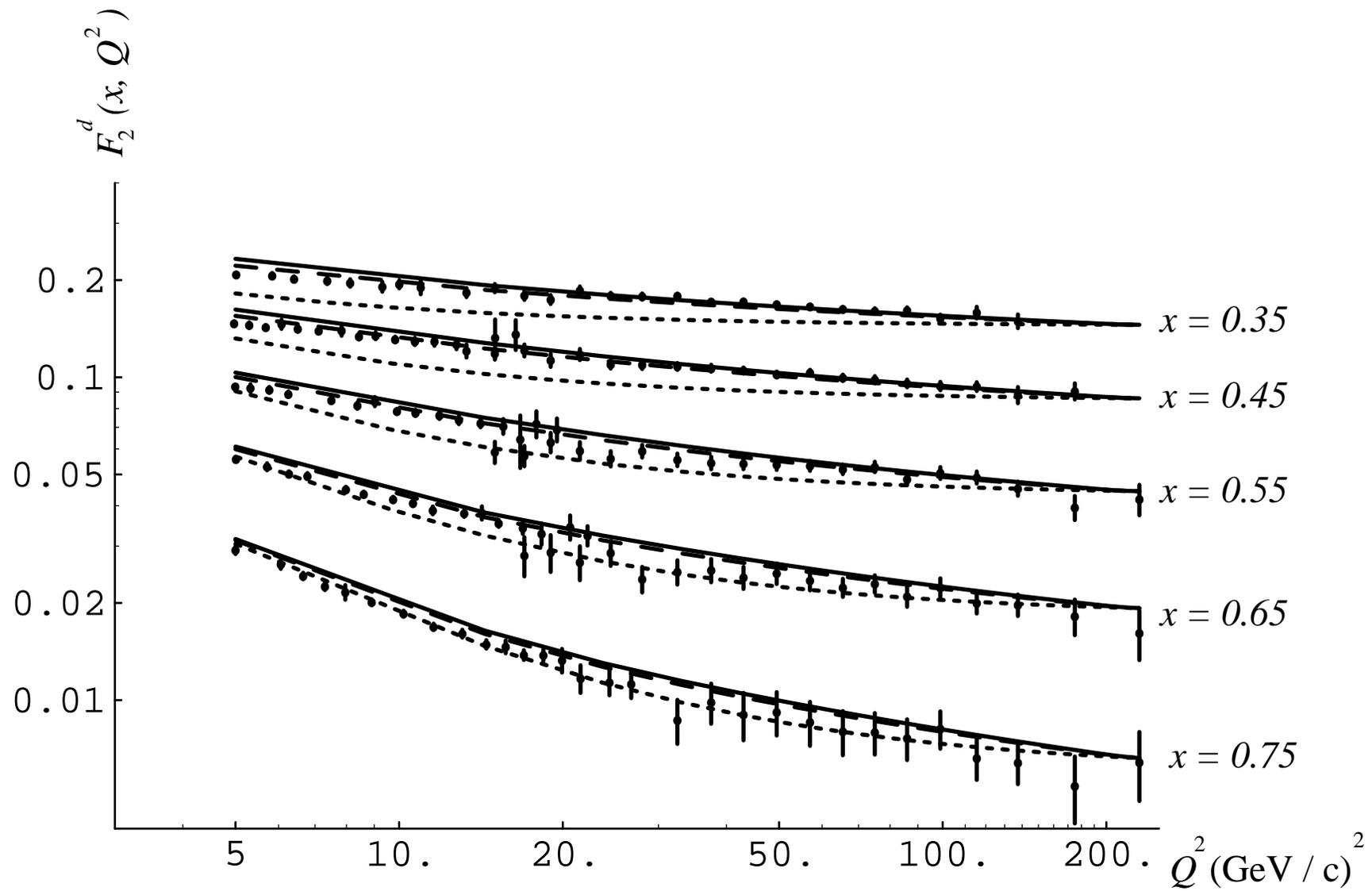
\noindent {\bf Fig 5.} {As in Fig. 4 for the deuteron structure function
$F_2^d(x,Q^2)$.}



\begin{references}
\bibitem{nacht1} O. Nachtmann, Nucl. Phys. B{\bf 38}, 397 (1972).

\bibitem{greenb}  O.W. Greenberg, Phys. Rev. D{\bf 47}, 331 (1993);
S.A. Gurvitz and A.S. Rinat, Phys. Rev. C{\bf 47}, 2901 (1993).

\bibitem{gur}  S.A. Gurvitz, Phys. Rev. D{\bf 52}, 1433 (1995).

\bibitem{slac1} L.W. Whitlow, Ph.D. Thesis, Stanford University, 1990,          SLAC-REPORT-357 (1990). 

\bibitem{slac2}  L.W. Whitlow {\em et al.},  Phys. Lett. B{\bf 282}, 
475 (1992).

\bibitem{slac3} S.E. Rock {\em et al.}, Phys. Rev. D{\bf 46}, 24 (1992).

\bibitem{slac4} P.E. Bosted {\em et al.}, Phys. Rev. D{\bf 49}, 3091 (1994).

\bibitem{bcd1} BCDMS Collab., A.C. Benvenuti {\em et al.}, 
Phys. Lett. B{\bf 223}, 485 (1989); Phys. Lett. B{\bf 237}, 592 (1989).

\bibitem{nmc1} NMC Collab., P. Amaudruz {\em et al.}, Phys. Lett. B{\bf 295}, 
159 (1992).

\bibitem{mel}  W. Melnitchouk, A.W. Schreiber and A.W. Thomas, 
Phys. Lett. B{\bf 335}, 11 (1994).

\bibitem{not1} Similar small binding and Fermi motion effects 
in the deuteron structure function were also found in a recent  
phenomenological analysis of J. Gomez {\em et al.}, 
Phys. Rev. D{\bf 49}, 4348 (1994). 

\bibitem{nlo1} R.G.Roberts, {\it The structure of the proton}, 
(Cambridge Univ. Press, Cambridge, 1990);
E.G. Floratos, C. Kounnas and R. Lacaze, Nucl. Phys. {\bf B192}, 417 (1981);
A. J. Buras, Rev. Mod. Phys. {\bf 50}, 199 (1980).

\bibitem{nmcfit} NMC Collab., M. Arneodo {\em et al.}, Phys. Lett. B{\bf 364}, 
107 (1995).

\bibitem {cteq95} CTEQ Collab., H.L. Lai {\em et al.}, Phys. Rev. D{\bf 51}, 
4763 (1995).

\bibitem{nlo2} M. Traini, A. Zambarda and V. Vento, Mod. Phys. Lett. 
{\bf 10}, 1235 (1995); M. Traini, V. Vento and A. Zambarda, to be published.

\bibitem{thesis}A. Zambarda, Thesis, Trento 1994, unpublished;
A. Mair, Thesis, Trento 1996, unpublished.

\end{references}
\end{document}